\documentclass[english,prd]{revtex4}
\usepackage[]{fontenc}
\usepackage[latin1]{inputenc}

\makeatletter

\newcommand{\noun}[1]{\textsc{#1}}

\usepackage{babel}
\makeatother
\begin{document}

\title{$\nu$ masses in a SUSY SO(10) theory with spontaneous CP violation}

\author{K. Bora}

\email{kalpana.bora@gmail.com}

\affiliation{Physics Department, Gauhati University, Guwahati 781 014, India}

\begin{abstract}
We propose a possibility of spontaneous CP-violation (SCPV) at high
scale in a SUSY SO(10) theory. The model is L-R symmetric SUSY SO(10)
with \textbf{\noun{10}} and \textbf{126} dimensional Higgs generating
fermion masses, and the CP phase is generated through complex VEV
of B-L breaking \textbf{126} Higgs . The model can have potential
application in explaining $\nu$ masses and leptogenesis as well.
\end{abstract}
\maketitle

\section{Introduction}

CP violation (CPV) was discovered experimentally about four decades
ago, but its origin still remains one of the fundamental open question
in particle physics. CPV is directly observed only in the decays of
the Kaons and B-mesons. At least four aspects of CPV are observed
in nature --- CPV in quark sector {[}1,2], in lepton sector, generation
of BAU (baryon asymmetry of the Universe), and strong and SUSY CP
problems. In this paper, we address its manifestation in the first
three cases only.

Within the premises of the standard model (SM) model, neutrinos are
mass-less, so there is no mixing and no CPV in the neutrino sector.
However, the existence of $\nu$ masses is a well established fact
now, and hence any theory which can explain them would also imply
CPV in leptonic (CPV-L) sector in principle. The latter might be detected
in future experiments to be performed at neutrino factories. Indirect
evidence for the CPV-L may also be provided by forthcoming experiments
on $\nu$-less $\beta\beta$ decay (Majorana phases). Existence of
matter dominated Universe is another evidence of CPV, but it has been
established that within the SM, it is not possible to generate observed
BAU, partly due to smallness of CPV in SM (CKM phase). This provides
motivation for considering new sources of CPV beyond the SM-CKM mechanism
(e.g. through CPV from SUSY breaking sector). In gauge theories, there
are two possibilities of CPV --- explicit (hard) CP breaking at the
Lagrangian level through complex Yukawa couplings (as in SM), or spontaneous
(soft ) CP breaking by the vacuum via complex VEV of the Higgs (SCPV).

SCPV at higher scales seems to be an interesting proposition to explain
the origin of CPV in nature since all the couplings of the Lagrangian
are real due to CP invariance at the Lagrangian level. CP is broken
only through phases in VEV of the Higgs {[}3,4]. There has been significant
amount of work in literature addressing the question of SCPV, an incomplete
list is given as {[}5-10]. It has been a known difficulty with SUSY
theories that they cannot generate CP breaking spontaneously. This
is because they lead to a real CKM matrix. A recent analysis {[}11]
of the present experimental data provides clear evidence for a complex
CKM matrix. These experimental findings {[}12] of the angle $\gamma$
inspires to ask the question if we can have a SUSY extension of the
SM with SCPV and a complex CKM matrix. In fact some work has already
been done in this line {[}6-9]. There, they introduce extra vector-like
quark which mixes with standard quarks and leads to a non-trivial
phase in $3\times3$ CKM matrix {[}7], VEV of \textbf{126} Higgs in
SUSY SO(10) theory is complex {[}6], or add extra Higgs {[}8], extension
of SM with a $\mbox{SU}(2)_{L}$ singlet quark and a singlet Higgs
field {[}10] , has been considered.

\section{Motivation for the present work}

In the present work, we have attempted to find a possible model to
generate a CP phase spontaneously in L-R symmetric SUSY SO(10) theory,
in particular in context of generating neutrino masses and mixings.
In the framework of SO(10) GUT SCPV was first discussed in {[}13].
In the present model, B-L symmetry is broken by a \textbf{126} dim
Higgs, which also contributes to fermion masses along with a \textbf{10}
dim Higgs {[}14,15]. This theory seems to be too attractive to generate
small $\nu$ masses --- it has a right handed Majorana neutrino (RHMN)
to implement see-saw mechanism, naturally contains B-L symmetry needed
to keep the RHMN below the Planck scale, provides a group theoretical
explanation of why neutrinos are Majorana particles, has automatic
R-parity conservation which leads to natural conservation of baryon
and lepton number symmetry prior to symmetry breaking, provides a
simple mechanism for explaining origin of matter in the Universe etc.
It has been shown that type-I see-saw predictions of this model are
in contradiction with experiments {[}14,16]. Then, type-II see-saw
{[}17] for neutrino masses {[}18] was suggested to explain the data.
In {[}19] $b-\tau$ unification was used to explain the $\nu$ masses
and mixings. In {[}20], CPV was introduced through complex Yukawa
couplings (of course this list is incomplete!), and it was found that
compatibility with $\nu$ data requires CKM phase to be outside the
first quadrant (whereas the SM CKM phase is in first quadrant). It
implies that to understand CPV in this minimal SO(10) model, one must
have a non-CKM source for CPV. So it would be interesting to see how
CPV can be generated in this model to explain $\nu$ masses and mixings,
along with generation of BAU through leptogenesis with the minimal
modification. Attempts have already been made in this direction {[}21,22],
but they included \textbf{120}-dim Higgs and CP is violated through
complex Yukawa couplings.

In the present work, we propose a different scenario. We show that
if assume SCPV at higher scales in minimal SUSY SO(10) theory with
complex VEV for B-L breaking \textbf{126} Higgs, one can't have a
nontrivial CPV phase. So we propose that if we include two \textbf{126}
Higgs, one with a real VEV and the other with a complex VEV, one can
have a nontrivial value of CPV phase by some fine-tuning in the Higgs
coupling constants of the Lagrangian. Now since the theory has $SU(4)_C$
symmetry at higher scales, the CKM phase in the baryonic quark mass
matrix will be related to CP phases in the leptonic sector as well.
And since heavy RH Majorana neutrino mass matrix will be complex (due
to complex VEV of \textbf{126} Higgs) the model has the potential
to explain BAU also. But, we would like to stress that we are not
attempting to comment on other issues such as strong CP, SUSY CP problems
etc, which can be solved may be by imposing some additional symmetries
on the Lagrangian, or via some other mechanism. This lies beyond the
scope of this paper.

Now, we would like to present the distinguishing features and novelties
of our work, as compared to some of the recent works in this line:

1. In {[}21,22], CP is introduced through complex Yukawa couplings,
whereas we use SCPV.

2. In {[}9], SUSY SO(10) $\rightarrow$ SM via intermediate SU(5),
while here we have used SU(2)$_L \times $SU(2)$_R \times $SU(4)$_C$
as the intermediate symmetry, although the breaking is single step
(i.e. M$_U$=M$_R${[}15]). 

3.In {[}8], they add extra Higgs/Higgs+fermions to the SM, while here
we have added extra \textbf{126} Higgs to the SUSY SO(10) theory,
with VEV of one of the \textbf{126} as real, and that of the other
as complex. At the same time, we have applied it for specific model
building purpose (for neutrino masses).

\section{The Model}

We consider the SUSY SO(10) theory, with \textbf{45}(A)+\textbf{54(S)}
dim Higgs field breaking SO(10) down to the L-R symmetric group SU(2)$_L \times $SU(2)$_R \times $U(1)$_{(B-L)}\times$ SU(4)$_C$
(G$_{2213}$), and the minimal Higgs set $10+126+\bar{126}$ that
couple to matter and also break the G$_{2213}$ group to G$_{31}$
(SU(3)$_C\times$U(1)$_{em}$) {[}15] (these details are well established,
but for the sake of completeness, we shall review them briefly here).
The Majorana mass of heavy RH neutrino owes its origin to the breaking
of local $B-L$ symmetry, therefore $M_R\sim M_{\rm seesaw} \sim M_{B-L}$
{[}23]. Local $B-L$ symmetry provides a natural way to understand
smallness of RH neutrino mass compared to $M_{\rm Pl}$. With G$_{2213}$
one can understand parity violation in nature. \textbf{126} ($\Delta)$
leaves R-parity as an exact symmetry, and explains why neutralino
can act as stable dark matter candidate {[}24]. In a generic SO(10)
with \textbf{126} getting VEV, one gets two contributions to $\nu$
masses - type-I see-saw and type-II see-saw (from induced VEV of the
triplet Higgs). The superpotential also contains Planck scale induced
non-renormalisable terms (more than dim-3), to give induced VEV to
triplet of \textbf{126} (for type-II see-saw). If these are not included,
\textbf{210} Higgs is needed, but getting DTS is not very simple {[}25]
here. The superpotential of the theory contains three parts,

\begin{equation}
W=W_{f}+W_{s}+W_{p},W_{f}=h_{ab}\Psi_{a}\Psi_{b}H+f_{ab}\Psi_{a}\Psi_{b}\bar{\Delta},\end{equation}
where $W_{f}$ generates mass of matter, with (2,2,1) of \textbf{10}
dim H and (2,2,15) of \textbf{126} dim Higgs acquiring VEV, $\Psi$
is the 16-dim spinor (matter field) of SO(10). The $W_{s}$ contains
scalar Higgs contribution, and is

\begin{eqnarray}
W_{s} & = & (\mu_{H}+\lambda S)HH+\mu_{s}S^{2}+\lambda_{s}S^{3}+\mu_{A}A^{2}+\mu_{\Delta}\Delta\bar{\Delta}+\lambda_{\Delta}\Delta A\bar{\Delta}\\
 &  & +\lambda'_{s}(S\Delta\Delta+S\bar{\Delta}\bar{\Delta})+\lambda_{A}SA^{2}.\nonumber \end{eqnarray}
The $W_{P}$ is Planck-scale induced part of the superpotential

\begin{equation}
W_{P}=\frac{\sqrt{8\pi}}{M_{Pl}}\lambda_{P}\Delta A^{2}H.\end{equation}
Now, from the superpotential, the F-term (Higgs part) of the potential
can be constructed as $V=\sum_{i}\left|\frac{\partial W}{\partial\sigma}\right|^{2}$,
where $\sigma$s are the Higgs scalars,\begin{equation}
V=\left|\frac{\partial W}{\partial\Delta}\right|^{2}+\left|\frac{\partial W}{\partial\bar{\Delta}}\right|^{2}+\left|\frac{\partial W}{\partial S}\right|^{2}+\left|\frac{\partial W}{\partial A}\right|^{2}+D-{\rm term},\end{equation}
and it is easy to see that $\left\langle A\right\rangle \Delta\Delta\bar{\Delta}H$
term from $\left|\frac{\partial W}{\partial A}\right|^{2}$ will contribute
induced VEV to $\Sigma(2,2,15)$ of 126 Higgs, to correct mass relations
of fermions, while $\Delta\Delta HH$ from $\left|\frac{\partial W}{\partial S}\right|^{2}$
term will contribute induced VEV to triplet $\Delta_{L}(3,1,\bar{10})$
for type-II see-saw mechanism. 

Next, we shall consider how the breaking of higher symmetries is realized
through VEVs of Higgs along different directions. In a SUSY theory,
the ground state should have zero energy, so both F-flatness and D-flatness
conditions must be satisfied. The latter is ensured by the presence
of both $\Delta$ and $\bar{\Delta}$. The F-flatness conditions,
with the scalars acquiring VEVs as follows are\begin{eqnarray}
\left\langle S\right\rangle  & = & {\rm diag}(k,k,k,k,k,k,k',k',k',k'),\left\langle A\right\rangle =i\tau_{2}\times{\rm diag}(b,b,b,c,c),\\
\left\langle \Delta\right\rangle  & = & v_{R}e^{i\delta},\,\,\left\langle \bar{\Delta}\right\rangle =v_{R}e^{-i\delta},\\
F_{s}:\frac{\partial W}{\partial k} & = & 2\mu_{s}k+3\lambda_{s}k^{2}+\lambda'_{s}(x_{0}v_{R}^{2}e^{2i\delta}+x_{0}v_{R}^{2}e^{-2i\delta})-\lambda_{A}b^{2}=0,\\
\frac{\partial W}{\partial k'} & = & 2\mu_{s}k'+3\lambda_{s}k'^{2}+\lambda'_{s}(v_{R}^{2}e^{2i\delta}+v_{R}^{2}e^{-2i\delta})-\lambda_{A}c^{2}=0,\\
F_{A}:\frac{\partial W}{\partial b} & = & -2b\mu_{A}+\lambda_{\Delta}x_{0}v_{R}^{2}-2b\lambda_{A}k=0,\\
\frac{\partial W}{\partial c} & = & -2c\mu_{A}+\lambda_{\Delta}v_{R}^{2}-2c\lambda_{A}k'=0,\\
F_{\Delta} & = & \mu_{\Delta}v_{R}e^{-i\delta}+\lambda_{\Delta}(x_{0}b+c)v_{R}e^{-i\delta}+2\lambda'_{s}(y_{o}k+k')v_{R}e^{i\delta}=0,\\
F_{\bar{\Delta}} & = & \mu_{\Delta}v_{R}e^{i\delta}+\lambda_{\Delta}(x_{0}b+c)v_{R}e^{i\delta}+2\lambda'_{s}(y_{o}k+k')v_{R}e^{-i\delta}=0,\end{eqnarray}
where $x_{0}$ and $y_{0}$ are appropriate C-G coefficients, due
to involvements of different groups. These constraints must give a
non-trivial solution for the CPV phase $\delta$. The $F_{\Delta}$
and $F_{\bar{\Delta}}$ constraints can be written as 

\begin{eqnarray}
(A+B)\cos\delta+i(A-B)\sin\delta & = & 0,\\
(A+B)\cos\delta+i(B-A)\sin\delta & = & 0,\end{eqnarray}
where constants $A$ and $B$ involve various Higgs couplings and
VEVs etc. It is easy to see that these equations give only the trivial
solutions $\delta=0$ and $\delta=\pi/2.$ These values of $\delta$
have also to be satisfied simultaneously by the $F_{s}$ constraints,\begin{eqnarray}
F_{s_{k}}:\cos2\delta & = & \frac{\lambda_{A}b^{2}-2\mu_{s}k+3\lambda_{s}k^{2}}{2\lambda'_{s}x_{o}v_{R}^{2}},\\
F_{s_{k'}}:\cos2\delta & = & \frac{\lambda_{A}c^{2}-2\mu_{s}k'+3\lambda_{s}k'^{2}}{2\lambda'_{s}v_{R}^{2}}.\end{eqnarray}
Eqs. (13-16) are the new results of our present work, which implies
that one can not have a nontrivial value of the CPV phase in a L-R
symmetric minimal SUSY SO(10) theory, where CP has been broken spontaneously
at high scale by the complex VEV of \textbf{126} Higgs.

\subsection{New proposal}

To overcome this difficulty, therefore, we propose that in the model,
we have two \textbf{126} Higgs, $\Delta_{1}$ and $\Delta_{2}$, such
that one of them acquires a real VEV while the other one a complex
VEV,

\begin{equation}
\left\langle \Delta_{1}\right\rangle =v_{R}e^{-i\delta},\,\,\left\langle \Delta_{2}\right\rangle =\epsilon v_{R},\end{equation}
here $\epsilon$ is a fine tuning parameter, which can be adjusted
to get a desired nontrivial value of CPV phase at higher scales {[}see
Eqs.(15-17)]. Note that this is not possible in a theory with one
\textbf{126}, or with two \textbf{126}s with same VEVs (real or complex).
The terms of the Lagrangian involving products of the form $\Delta_{1}\Delta_{2}$
will help us get values of CPV phase other than 0 or $\pi/2$, through
the structure of Eqs. (13-16). The part of the new superpotential
generating fermion masses will look like,

\begin{equation}
W_{f}=h_{ab}\Psi_{a}\Psi_{b}+f_{1ab}\Psi_{a}\Psi_{b}\bar{\Delta}_{1}+f_{2ab}\Psi_{a}\Psi_{b}\bar{\Delta}_{2},\end{equation}
and accordingly, one can have new formulas for neutrino masses. Since
the VEV of a \textbf{126} is complex, the fermion mass matrices, the
CKM matrix and the heavy right handed Majorana mass matrix will be
complex.

\section{Conclusions}

To conclude, we have presented a novel mechanism of generating CP
violating phase spontaneously at higher scales in a L-R symmetric
SUSY SO(10) theory, which can be further applied in context of neutrino
masses and mixings, and leptogenesis. Eq. (17) is the new idea proposed
here for the first time, in this work, which together with Eqs. similar
to (13-16) can give a nontrivial CP violating phase (other than 0
or $\pi/2$) in the theory. We have shown this explicitly through
the F-flatness conditions. Of course further investigations, as far
as the applications and implications of this idea are concerned, are
needed, which can be taken up in future works.

\end{document}